\documentclass[aps,prl,twocolumn]{revtex4}
\usepackage{graphicx}

\begin{document}

\title{Kolmogorov scaling and intermittency in Rayleigh-Taylor turbulence}

\author{G. Boffetta$^{(1)}$, A. Mazzino$^{(2)}$, S. Musacchio$^{(1)}$, L. Vozella$^{(2)}$}
\affiliation{$^{(1)}$Dipartimento di Fisica Generale and INFN, 
Universit\`a di Torino, via P.Giuria 1, 10125 Torino (Italy)  \\
and CNR-ISAC, Sezione di Torino, corso Fiume 4, 10133 Torino (Italy) \\
$^{(2)}$Dipartimento di Fisica, Universit\`a di Genova, INFN and CNISM, 
via Dodecaneso 33, 16146 Genova (Italy)}

\date{\today}

\begin{abstract}
The Rayleigh--Taylor (RT) turbulence is investigated by means of 
high resolution numerical simulations.
The main question addressed here is on whether
RT phenomenology  can be considered as a manifestation of universality
of Navier--Stokes equations with respect to forcing mechanisms.
At a theoretical level the situation is far from being firmly established
and, indeed, contrasting predictions have been formulated.  
Our first aim here is to clarify the above controversy through a deep 
analysis of scaling behavior of relevant statistical observables. 
The effects of intermittency on the mean field scaling predictions
is also discussed.
\end{abstract}

\pacs{PACS?}
%\keywords{}

\maketitle

The Rayleigh--Taylor (RT) turbulence is a well-known buoyancy
induced fluid-mixing mechanism occurring in a variety of situations ranging
from geophysics (see, e.g., Ref.~\cite{mammatus06} in relation to
cloud formation) to astrophysics (in relation to thermonuclear reactions
in type-Ia supernovae \cite{zwrdb_aj05,cc_natphys06} and heating of
solar coronal \cite{imsy_nat05}) to technological related
problems, e.g., inertial confinement fusion (see Ref.~\cite{fujioka04}).

Despite the ubiquitous nature of RT turbulence, a consistent 
phenomenological theory has been proposed only recently 
\cite{chertkov_prl03}.
In three-dimensions, this theory predicts a Kolmogorov-Obukhov 
turbulent cascade in which temperature fluctuations are 
passively transported.
This scenario, which is partially supported
by numerical simulations \cite{cc_natphys06,vc_arxiv08},
has however been contrasted by an alternative picture which
rules out Kolmogorov phenomenology \cite{poujade_prl06}.

The goal of our work is twofold. From one hand we give stronger numerical
support to the phenomenological theory {\it \`a la} Kolmogorov in RT 
turbulence. On the other hand, we push the analogy with usual 
Navier-Stokes (NS) turbulence much further: we find that small scale
velocity fluctuations in RT turbulence develop intermittent statistics
analogous to NS turbulence.

%The goal of our work is twofold: from one hand we give stronger numerical
%support to the phenomenological theory {\it \`a la} Kolmogorov; on the
%other hand we push a step further the description of RT phenomenology
%to account for intermittency corrections. 

%%%%%%%%%%%%%%%%%%%%%%%%%%%%%%%%%%%%%%%%%%%%%%%%%%

We consider the 3D, incompressible (${\bf \nabla} \cdot {\bf v}=0$),
miscible Rayleigh-Taylor flow
in the Boussinesq approximation
\begin{eqnarray}
&& \partial_t {\bf v} + {\bf v} \cdot {\bf \nabla} {\bf v} = - {\bf \nabla} p
+ \nu \triangle {\bf v} + \beta {\bf g} T \label{eq:1} \\
&& \partial_t T + {\bf v} \cdot {\bf \nabla} T = \kappa \triangle T
\label{eq:2}
\end{eqnarray}
where $T$ is the temperature field, proportional to density
via the thermal expansion coefficient $\beta$,
$\nu$ the kinematic viscosity, $\kappa$ the molecular
diffusivity and ${\bf g}=(0,0,g)$ is the gravitational acceleration.

%------------------------------------------------------------------------
\begin{figure}[h!]
\centering
\begin{tabular}{cc}
\includegraphics[keepaspectratio,width=3.8cm]{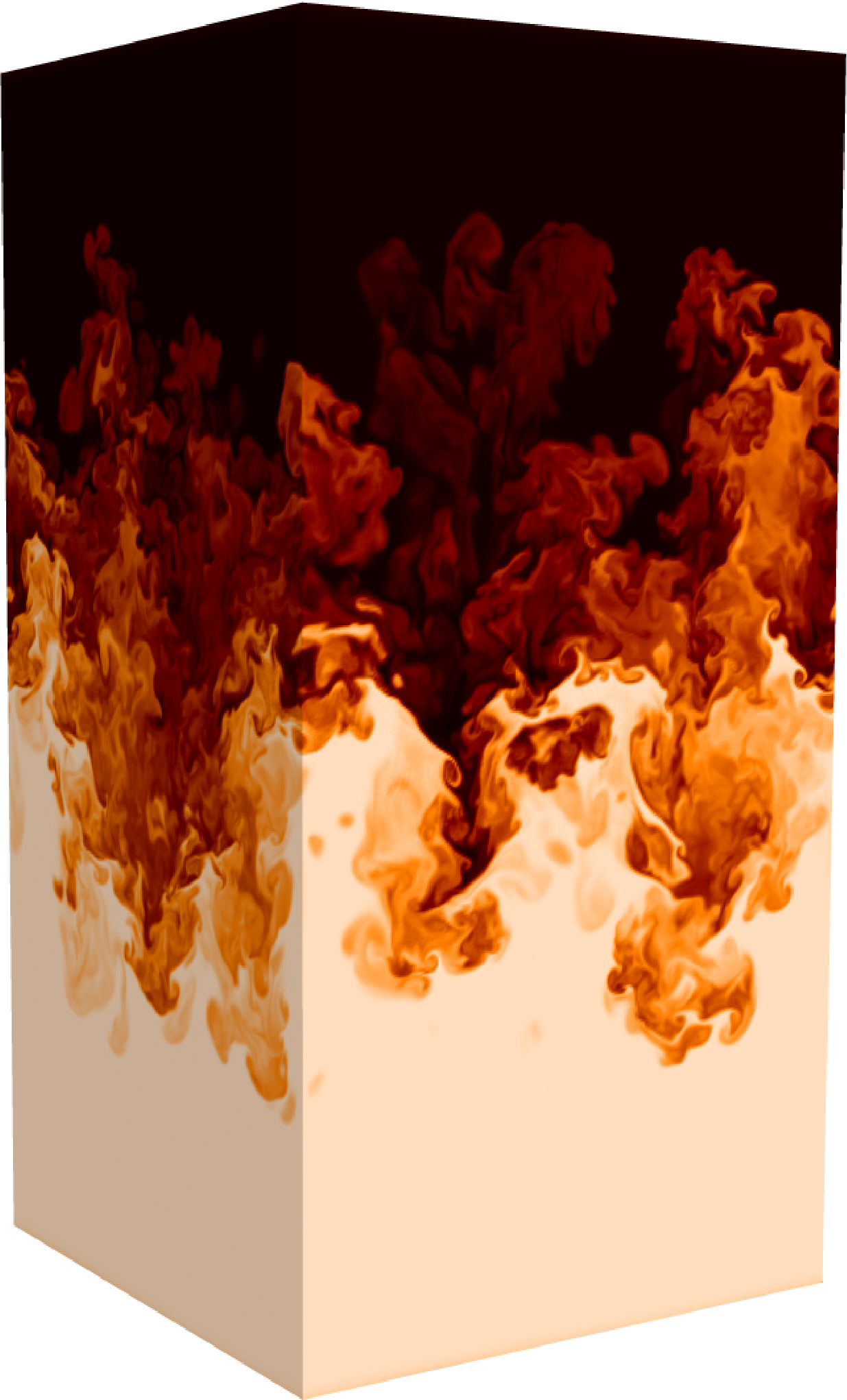} &
\includegraphics[clip=true,keepaspectratio,width=4.6cm]{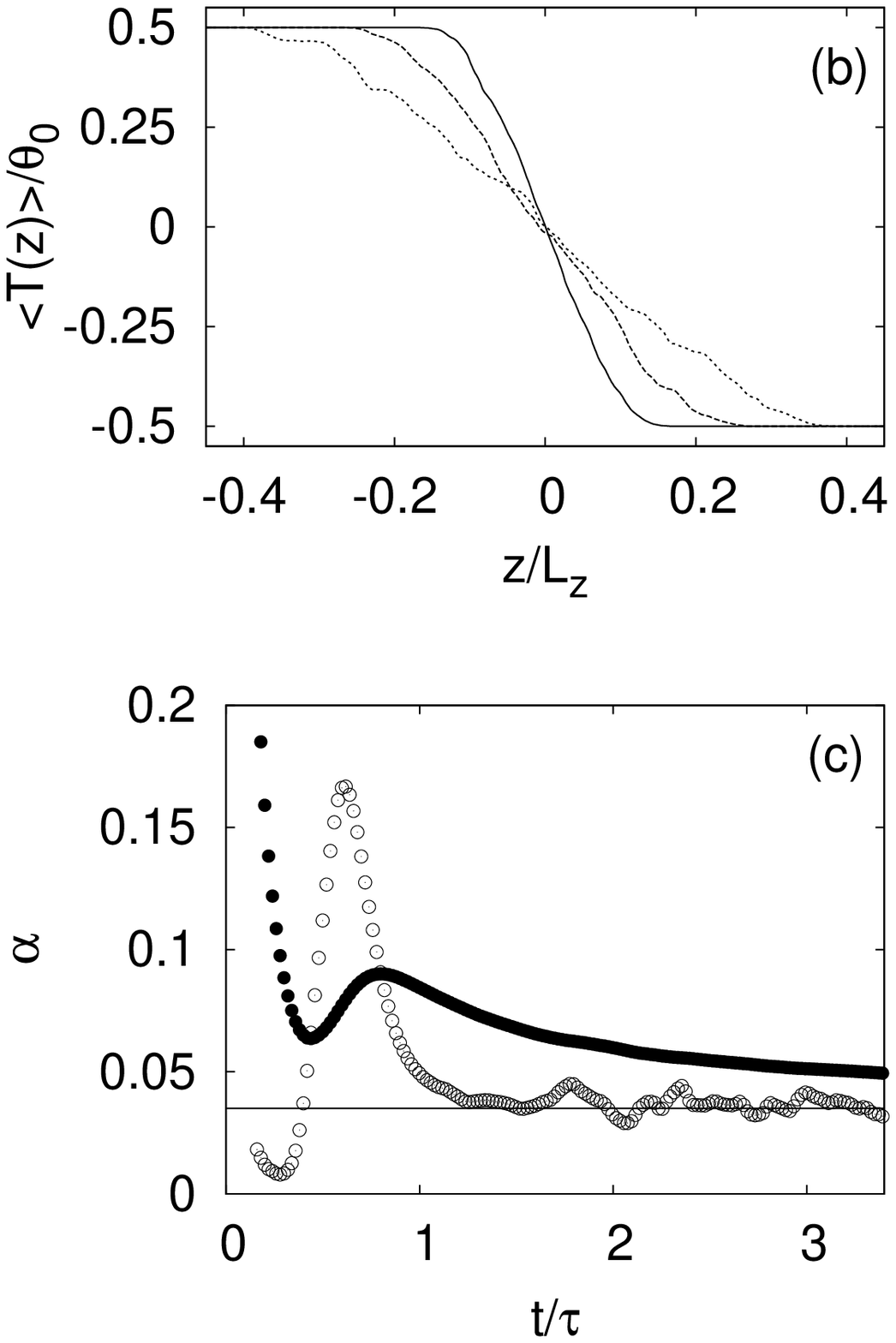} \\
\end{tabular}
\caption{(a) Snapshot of temperature field for Rayleigh-Taylor turbulence
at $t/\tau=2.6$. White (black) regions correspond to hot (cold)
fluid.
(b) Mean temperature profiles $\langle T(z) \rangle$ for times
$t/\tau=1.4$ (continuous), $t/\tau=2.0$ (dashed) and $t/\tau=2.6$ (dotted).
(c) Growth of the mixing layer thickness $h(t)$ defined as
the vertical range for which $|\langle T(z) \rangle| \le 0.98 \theta_0/2$
compensated with the dimensional prediction $A g t^2$ in order to get
the dimensionless coefficient $\alpha$.
Filled symbols: $\alpha = h/(A g t^2)$, open symbols:
$\alpha = \dot{h}^2/(4 A g h)$ \cite{cc_natphys06}.}
\label{fig1}
\end{figure}
%------------------------------------------------------------------------

At time $t=0$ the system is at rest with cooler (heavier) fluid placed above
the hotter (lighter) one. This corresponds to ${\bf v}({\bf x},0)=0$ and
to a step function for the initial temperature profile:
$T({\bf x},0)=-(\theta_0/2) \mbox{sgn}(z)$ where $\theta_0$
is the initial temperature jump which fixes the
Atwood number $A=(1/2) \beta \theta_0$.
The development of the instability leads to a mixing zone
of width $h$ which starts from the plane $z=0$ and is dimensionally
expected to grow in time according to
$h(t)= \alpha A g t^2$ \cite{dimonte_etal_pof04,cc_natphys06}.
Inside this mixing zone, turbulence develops in space and time. The
phenomenological theory \cite{chertkov_prl03} predicts for
velocity and temperature fluctuations the scaling laws
\begin{eqnarray}
\delta_r v(t) &\simeq & (A g)^{2/3} t^{1/3} r^{1/3}
\label{eq:3} \\
\delta_r T(t) &\simeq & \theta_0 (A g)^{-1/3} t^{-2/3} r^{1/3}
\label{eq:4}
\end{eqnarray}
The first relation represents Kolmogorov scaling with a time dependent
energy flux $\epsilon \simeq (A g)^2 t$.
>From the above scaling laws
one obtains that the buoyancy term $\beta g T$
becomes subleading at small scales in (\ref{eq:1}), consistently
with the assumption of passive transport of temperature fluctuations.

We integrate equations (\ref{eq:1}-\ref{eq:2})
by a standard $2/3$-dealiased pseudospectral method on a periodic
domain with uniform grid spacing, square basis $L_x=L_y$ and aspect
ratio $L_x/L_z=r$ with a resolution up to $512 \times 512 \times 2048$ 
($r=1/4$).
Time evolution is obtained by a second-order Runge-Kutta scheme
with explicit linear part. In all runs, $A g=0.25$,
$Pr=\nu/\kappa=1$, $\theta_0=1$. Viscosity is sufficiently large
to resolve small scales ($k_{max} \eta \simeq 1.2$ at final time).
In the results, scales and times are made dimensionless
with the box scale $L_z$ and the characteristic time
$\tau=(L_z/A g)^{1/2}$ \cite{dly_jfm99}.

Rayleigh-Taylor instability is seeded by perturbing the initial
condition with respect to the step profile. Two different perturbations
were implemented in order to check the independence of the turbulent
state from initial conditions. In the first case the interface $T=0$
is perturbed by a superposition of small amplitude waves in a
narrow range of wavenumber around the most unstable linear mode
\cite{rda_jfm05}. For the second set of simulations, we perturbed
the initial condition by ``diffusing'' the interface around $z=0$.
Specifically, we added a $10 \%$ of white noise to the value
of $T({\bf x},0)$ in a small layer of width $h_0$ around $z=0$.

Figure~\ref{fig1} shows a snapshot of the temperature field
for a simulation with $r=1/2$ at advanced time.
Large scale structures (plumes) identify the direction of gravity
and break the isotropy.
Nonetheless, we find that at small scales isotropy is almost completely
recovered: the ratio of vertical to horizontal rms velocity is
$v_z/v_x \simeq 1.8$ while for the gradients we have
$\partial_z v_z/\partial_x v_x\simeq 1.0$.
The horizontally averaged temperature $\langle T(z) \rangle$ follows
closely a linear profile within the mixing layer where, therefore,
the system recovers statistical homogeneity.

The analysis of the mixing layer width growth is also presented
in Fig.~\ref{fig1}. As shown by previous studies \cite{rc_jfm04,cc_natphys06},
the naive compensation of $h(t)$ with $A g t^2$ does not give
a precise estimation of the coefficient $\alpha$ because of the presence
of subleading terms which decay slowly in time. We have therefore
implemented the similarity method introduced in \cite{cc_natphys06}
which gives an almost constant value of $\alpha \simeq 0.038$ for
$t/\tau \ge 1.5$, consistent with previous studies
\cite{rc_jfm04,dimonte_etal_pof04}.

%------------------------------------------------------------------------
\begin{figure}[htb!]
\includegraphics[clip=true,keepaspectratio,width=8.0cm]{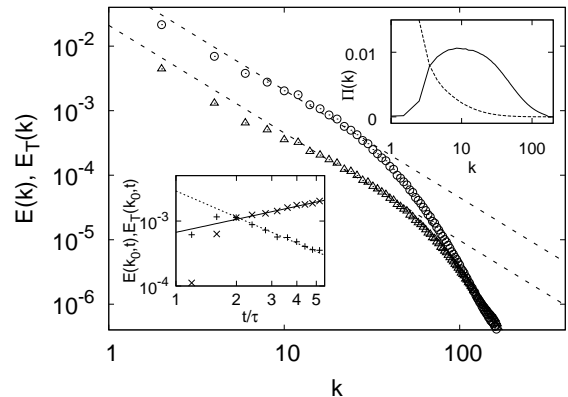}
\caption{Two-dimensional kinetic energy spectrum ($\circ$) and
temperature spectrum ($\triangle$) at time $t/\tau=2.6$
corresponding to $R_{\lambda}=245$. 
Spectra are computed by Fourier transforming
velocity and temperature fields on two-dimensional horizontal planes
and averaging over $z$ in the mixing layer.
Dashed lines represent Kolmogorov scaling $k^{-5/3}$.
Lower inset: evolution in time of the amplitude of kinetic energy ($\times$)
and temperature ($+$) spectra at fixed wavenumber $k_0=12$.
Lines represent the dimensional predictions $t^{2/3}$ (continuous)
and $t^{-4/3}$ (dashed) given by (\ref{eq:3}) and (\ref{eq:4}).
Upper inset: inertial (continuous) and buoyancy (dashed) contributions
to kinetic energy flux $\Pi(k)$ in Fourier space.}
\label{fig2}
\end{figure}
%------------------------------------------------------------------------

Figure~\ref{fig2} shows the kinetic energy $E(k)$ and temperature
$E_{T}(k)$ spectra within the similarity regime.
>From  (\ref{eq:3}) and (\ref{eq:4}), we expect the following
spatial-temporal
scaling of spectra: $E(k,t) \sim t^{2/3} k^{-5/3}$ and
$E_T(k,t) \sim t^{-4/3} k^{-5/3}$. Kolmogorov scaling $k^{-5/3}$
is evident for both velocity and temperature fluctuations.
Moreover, self-similar temporal evolution of spectra is well reproduced,
as shown in the lower inset. 
Also in Fig.~\ref{fig2} the two contributions to kinetic energy flux
in spectral space are shown. Buoyancy contribution, dominant at large
scale, becomes subleading at smaller scales, in agreement with the
Kolmogorov-Obukhov picture.
The above results, together with previous simulations
\cite{cc_natphys06,vc_arxiv08} and theoretical arguments
\cite{chertkov_prl03}, give a coherent
picture of RT turbulence as a Kolmogorov cascade of kinetic energy
forced by large scale temperature instability.

In the following we push this analogy one step ahead
by showing that small scale fluctuations in RT turbulence display
intermittency corrections typical of usual Navier--Stokes (NS) turbulence.
Intermittency in turbulence is a consequence of non-uniform transfer
of energy in the cascade which breaks down scale invariance. As a
consequence, scaling exponents deviates from mean field theory and
cannot be determined by dimensional arguments \cite{frisch_95}.
Several studies have been devoted to the intermittent statistics
in  NS turbulence, where the main issue concerns the
possible universality of anomalous scaling exponent with respect to the
forcing mechanisms and the large scale geometry of the flow.
While universality has been demonstrated for the simpler problem
of passive scalar transport, it is still an open issue for
nonlinear NS turbulence.
Therefore the key question is whether small scale statistics in
RT turbulence is equivalent to the statistics observed in
homogeneous, isotropic turbulence.

The simplest, and historically first, evidence of intermittency
is in the dependence of energy dissipation on Reynolds number
\cite{champagne_jfm78,va_pof80,ms_prl87}.
Classical statistical indicators are the flatness $K$ of
velocity derivatives \cite{va_pof80,ms_prl87} (corresponding to
$K \simeq \langle \epsilon^2 \rangle/\langle \epsilon \rangle^2$
in terms of energy dissipation), and the variance of the logarithm of
kinetic energy dissipation which is
expected to grow with Reynolds number as
\begin{equation}
\sigma_{\ln \epsilon} = a + (3 \mu/2) \ln R_{\lambda}
\label{eq:5}
\end{equation}
The exponent $\mu$ is the key ingredient for the log-normal model
of intermittency and its value is
determined experimentally \cite{va_pof80,sk_pof93}
and numerically \cite{ypld_pof06} to be
$\mu \simeq 0.25$.
More in general, moments of local energy dissipation
are expected to have a power-law dependence on $R_{\lambda}$
\begin{equation}
\langle \epsilon^p \rangle \simeq \langle \epsilon \rangle^p
R_{\lambda}^{\tau_p}
\label{eq:6}
\end{equation}
where the set of exponents $\tau_p$ can be predicted within the
multifractal model of turbulence \cite{frisch_95,br_pof02,bmv_jpa08} in
terms of the set of fractal dimensions $D(h)$.

%------------------------------------------------------------------------
\begin{figure}[htb!]
\includegraphics[clip=true,keepaspectratio,width=8.0cm]{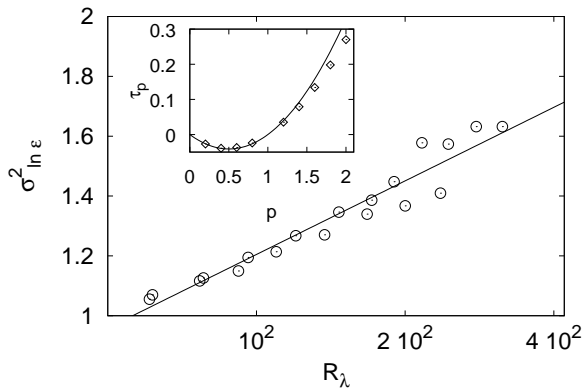}
\caption{Scaling of the variance of $\ln \epsilon$ on Reynolds
number defined as $R_{\lambda}=(v_z)_{rms} (\partial_z v_z)_{rms}/\nu$, obtained
from two realizations with white noise initial perturbation. The line
is the best fit corresponding to $\mu=0.24$ in (\ref{eq:5}).
Inset: scaling exponents of the moments of local dissipation
$\tau_p$ obtained from best fits according to (\ref{eq:6}).
The line represents the log-normal approximation $\tau_p=(3/4)\mu (p^2-p)$.}
\label{fig3}
\end{figure}
%------------------------------------------------------------------------

Because in RT turbulence the Reynolds number increases in time,
it provides a natural framework for a check of (\ref{eq:5}) and (\ref{eq:6}).
Figure~\ref{fig3} shows the dependence of the variance of $\ln \epsilon$
on $R_{\lambda}$ together with the first moments of energy dissipation.
Despite the limited range of $R_{\lambda}$, a clear scaling of
$\ln \epsilon$ is observable, even if with some fluctuations.
The best fit with (\ref{eq:5}) gives an exponent
$\mu \simeq 0.24$, very close to what observed in homogeneous,
isotropic turbulence \cite{ypld_pof06}.

Scaling exponents $\tau_p$ for the moments of dissipation
(\ref{eq:6}) are also shown in Fig.~\ref{fig3}. We were able
to compute moments up to $p=2$ with statistical significance.
Log-normal approximation, which is in general valid for $p \to 0$,
is found to be unsatisfactory for larger values of $p$. For
$p=2$, which corresponds to the flatness $K$ of velocity derivatives,
we find $\tau_2 \simeq 0.27$. 
This result is consistent with experiments at comparable 
Reynolds numbers \cite{va_pof80} which shows that $K \sim R_{\lambda}^{0.2}$
for $R_{\lambda}<200$ while an asymptotic exponent $\tau_2 \simeq 0.41$
is reached for $R_{\lambda}>10^3$ only.

%------------------------------------------------------------------------
\begin{figure}[htb!]
\includegraphics[clip=true,keepaspectratio,width=8.0cm]{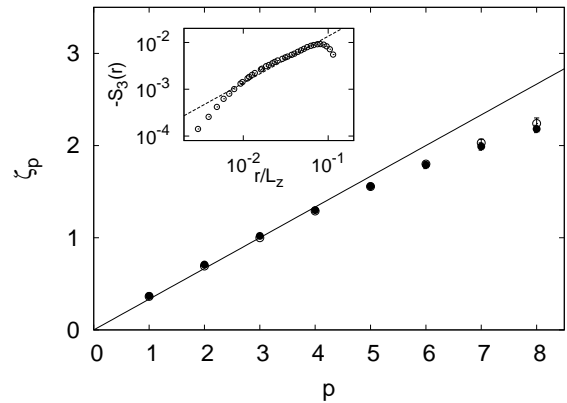}
\caption{Scaling exponents of isotropic longitudinal
velocity structure functions
$S_p(r)=\langle \left( \delta_r {\bf v} \cdot \hat{\bf r} \right)^p \rangle$
($\hat{\bf r}={\bf r}/r$) for the late stage of RT
turbulence (open circle). Exponents are computed by compensation of
$S_p(r)$ with $S_3(r)$, according to the ESS procedure
\cite{bctbms_pre93} averaging inside the mixing layer
and on all directions.
Filled circles: scaling exponents from simulations
of homogeneous, isotropic turbulence at $R_{\lambda}=381$
\cite{gfn_pof02}. Line represents dimensional prediction
$\zeta_p=p/3$.
Inset: third-order isotropic longitudinal structure function
$S_3(r)$. The line represents Kolmogorov's four-fifth law
$S_3(r)=-4/5 \epsilon r$.}
\label{fig4}
\end{figure}
%------------------------------------------------------------------------

In NS turbulence intermittency is also observed in the inertial range
of scales as deviations of velocity structure functions
$S_p(r)=\langle (\delta_r v)^p \rangle$ from the dimensional
prediction (\ref{eq:3}) which corresponds to
$S_p(r) \simeq r^{p/3}$ \cite{frisch_95}.
Anomalous scaling is observed, which corresponds to scaling laws
$S_p(r) \simeq r^{\zeta_p}$ with a set of exponents $\zeta_p \ne p/3$.
We remind that constancy of energy flux in the inertial range implies
$\zeta_3=1$ independently on intermittency, as required by the Kolmogorov's
``four-fifths'' law $S_3(r)=-4/5 \epsilon r$ \cite{frisch_95},
which is indeed observed in our simulations (see inset of Fig.~\ref{fig4}).
Figure~\ref{fig4} shows the first longitudinal scaling exponents computed
from our simulations exploiting the extended self-similarity procedure
which allows for a precise determination of the exponents at moderate
Reynolds numbers \cite{bctbms_pre93}. A deviation from
dimensional prediction $\zeta_p=p/3$ is clearly observable for
higher moments. Fig.~\ref{fig4} also shows
the scaling exponents obtained from a homogeneous, isotropic
simulation of NS equations at a comparable $R_{\lambda}$ \cite{gfn_pof02}.
The two sets agree within the error bars, this gives further quantitative
evidence in favor of the equivalence between RT turbulence and NS
turbulence in three dimensions.

We end by discussing the behavior of turbulent heat flux and rms
velocity fluctuations as a function of the mean temperature gradient. 
In terms of dimensionless variables, these quantities are represented
respectively by the Nusselt number
$Nu=1+\langle v_z T \rangle L/(\kappa \theta_0)$, the Reynolds numbers
$Re=v_{rms} L/\nu$ and Rayleigh number
$Ra=A g L^3/(\nu \kappa)$. The relations between these
quantities has been object of many experimental and numerical
studies in past years, mainly in the context of Rayleigh-B$\acute{e}$nard
turbulent convection
\cite{siggia_arfm94,gl_jfm00,nssd_nat00,na_prl03,lt_prl03,cmv_prl06}.
Experiments have reported both simple scaling laws $Nu \sim Ra^{\beta}$
with exponent $\beta$ scattered around $\beta=0.3$
\cite{gsns_nat99,nssd_nat00} and, more complicated behavior
\cite{xba_prl00,na_prl03} partially in agreement with a
phenomenological theory \cite{gl_jfm00}.
However, in the limit of very large $Ra$, Kraichnan \cite{kraichnan_pof62}
predicted an asymptotic scaling $Nu \sim Ra^{1/2}$ now called
the ultimate state of thermal convection. This regime is expected
to hold when thermal and kinetic boundary layers become irrelevant,
and indeed has been observed in numerical simulation of thermal
convection at moderate $Ra$ when boundaries are artificially 
removed \cite{lt_prl03}.
It is therefore natural to expect that the ultimate state scaling
arises in RT convection where boundaries are absent.

The ultimate state relations can formally be obtained
from kinetic energy and temperature balance equations
\cite{gl_jfm00}. In the context of RT turbulence,
they are a simple consequence of the dimensional scaling of
the mixing length
$L \equiv h \simeq A g t^2$ and of the rms velocity
$v_{rms}\simeq A g t$. Inserting in the
definition of the dimensionless numbers one obtains
\begin{equation}
Nu \sim Pr^{1/2} Ra^{1/2} \, , \mbox{\hspace{1cm}}
Re \sim Pr^{-1/2} Ra^{1/2}
\label{eq:7}
\end{equation}
where $Pr=\nu/\kappa$.

%------------------------------------------------------------------------
\begin{figure}[ht!]
\includegraphics[clip=true,keepaspectratio,width=7.0cm]{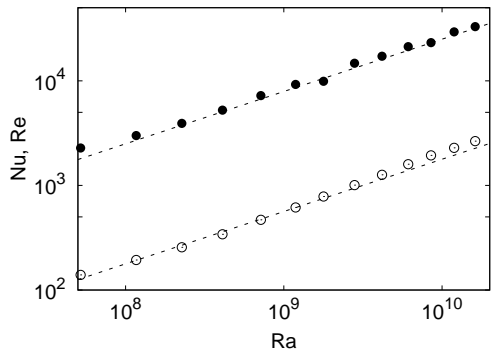}
\caption{The scaling of Nusselt number (open circles) and Reynolds
number (solid circles) as functions of Rayleigh number. Lines represent
the ultimate state predictions (\ref{eq:7}).}
\label{fig5}
\end{figure}
%------------------------------------------------------------------------

We remark that the above relations are independent on the statistics
of the inertial range and on the presence of intermittency as
they involve large scale quantities only.
Our numerical results, shown in Fig.~\ref{fig5}, confirms the 
ultimate state scaling (\ref{eq:7}). 
The same behavior has been predicted and
observed for two-dimensional RT  simulations, where
temperature fluctuations are not passive and Bolgiano scaling is
observed in the inertial range \cite{cmv_prl06}. 
The elusive Kraichnan scaling in thermal convection
finds its natural manifestation in Rayleigh-Taylor turbulence,
which turns out to be an excellent
setup for experimental studies in this direction.

\bibliography{biblio}{}

\end{document}